\documentstyle[psfig,multirow]{mn} 
\title[X-ray selected AGNs: The Clustering Properties] 
{X-ray selected AGNs: The Clustering Properties} 
\author[E.Ghaffarsedeh{\it et al.}]
{E.Ghaffarsedeh$^{1}$, J.Quenby$^{1}$\\  
$^1$Astrophysics Group, Blackett Laboratory, 
Imperial College, Prince Consort Rd, London SW7 2BZ, UK}

\begin{document}

\maketitle

\begin{abstract}

Clustering properties of two samples of X-ray selected AGNs are compared with IR/Optical
galactic samples. These are the Einstein Observatory Extended Medium 
Sensitivity Survey: EMSS (Maccacaro {it et al.} 1991), with flux limit of 
$> 3.25$(10$^{-14}$) erg cm$^{-2}$ sec$^{-1}$ (0.5-2.0 keV), redshifts up to 2.83 and a 
sky coverage of 778 deg$^{2}$; and an all sky sample of 654 ROSAT detected radio quasars 
(Brinkmann {\it et al.} 1997), with a uniform flux limit of a few 
(10$^{-13}$) erg cm$^{-2}$ sec$^{-1}$ (0.1-2.4 keV) and redshifts up to 3.886.
The cosmological dipole moment is calculated for these samples, yielding the values 
of $b \Omega_o^{-0.6}$ (where b is the bias parameter, and $\Omega_o$ is the
present value of the cosmological density parameter). We find this factor to be 
$0.68 \pm 0.22$ for the EMSS sample, and $1.02 \pm 0.014$ for the quasars sample. 
Comparing these values with that of the IRAS 1.2 Jy sample $(1.95 \pm 0.096)$, 
suggests that these AGNs are less clustered than the IRAS galaxies. For each sample there 
is a depth at which the amplitude of the dipole growth curve saturates. This convergence 
radius $(R_{conv})$ is found to be $\sim 300-400$ h$^{-1}$ Mpc for the EMSS sample and 
$\sim 450-500$ h$^{-1}$ Mpc for the ROSAT detected quasars sample, which suggests that X-ray AGNs
have a much deeper contribution to the Local Group (LG) motion compared to the galaxy
and cluster samples. Also the maximum amplitudes at saturation are found to be 
$~ 405 \pm 103$ km/sec for the EMSS AGNs and $~ 617 \pm 8$ km/sec for the quasar sample, 
which suggests that the lower the flux limit of the sample (ie. the more sensitive the sample), 
the lower the maximum amplitude at the region of saturation, therefore the less clustered the 
sample. Using the properties of the angular and spatial correlation functions for these samples,
we find: $b_{IRAS}/b_{quasars} = 3.61 \pm 2.7$; $b_{IRAS}/b_{EMSS} = 8.46 \pm 6.4$ and 
$b_{quasars} / b_{EMSS} = 2.35 \pm 0.2$. Using Chi-squared minimisation, we have fitted the 
angular correlation functions using the standard power law model of the form: 
$W(\theta)= A \theta^{-\delta}$, and the spatial correlation functions using the power law 
model of the form: $\xi(r) =(r/r_{0})^{-\gamma}$. For the angular correlation 
functions, we find $A \sim0.6$\ for the AGN auto-correlations and $A \sim0.2-0.3$\
for the AGN-IRAS galaxy cross-correlations. Also $\delta$ is found to be $~ 0.8-1.1$
. In the spatial correlation analysis, we find $r_{o}\sim3.5-5.1$ $h^{-1}Mpc$ and 
$\gamma\sim1.4-1.7$\ . The bias factor resulting from the $IR/Optical$ galaxies may not be 
the true representative of this parameter, due to the local nature of those samples, 
whereas AGN samples are deep/far enough to give a more accurate value of this parameter.

\end{abstract} 
 
 
\section{Introduction}

A way of investigating the large scale clustering properties of extragalactic objects, 
is to analyse their cosmological dipole moment. The anisotropy of the Cosmic Microwave 
Background $(CMB)$ radiation and the solar motion relative to the Local Group of galaxies 
$(LG)$ implies that the LG has a peculiar velocity of $\sim600$ km/sec towards 
(l,b) = (277$^{o}$,30$^{o}$) (cf. Rowan-Robinson, Lynden-Bell, 1988). 
According to the linear perturbation theory (Peebles, 1980), the peculiar motion of 
the LG is in the same direction as the gravitational acceleration arising from the 
distribution of mass outside of the LG. Attempts to trace this mass distribution responsible 
for the gravitational acceleration associated with the peculiar velocity of the LG have
been made using the IRAS galaxies (Yahil, Walker \& Rowan-Robinson,
1986; etc.), 
Optical galaxies (Lahav, 1987; etc.), X-ray clusters (Lahav et al., 1989) and 
X-ray AGNs (Miyaji \& Bolt, 1990).
 
\noindent The spatial extend of the distribution of 
mass inhomogeneities that cause the LG motion has been a matter of debate. 
The most probable cause for this
motion as well as for the observed peculiar motions of other galaxies and 
clusters is gravitational instability. This is supported by the fact that the 
gravitational dipole (acceleration) of many different samples of extragalactic 
mass tracers is well aligned with the general direction of the CMB dipole 
(Rowan-Robinson et al., 1990; Strauss et al. 1992; etc.).
 
\noindent However what still seems to be under discussion is the depth from which density
fluctuations contribute to the gravitational field that shapes the LG motion.
The largest such depth, is defined by the dipole convergence depth, $R_{conv}$,
which is that depth where the true gravitational acceleration converges to its 
final value. The apparent value of $R_{conv}$ differs from sample to sample, 
normally in the range 40 to 100 h$^{-1}$ Mpc (for galaxy samples), with a 
strong 
dependence on the sample's characteristic depth. Cluster samples, such as
Optical $Abell / ACO$ cluster sample is volume limited out to large
enough depth ($\sim 240 h^{-1}$ Mpc) to allow a more reliable
determination of $R_{conv}$ which was 
found to be $\sim 160 h^{-1}$ Mpc (Scaramella et al., 1991; Plionis \&
Valdernini, 1991; 
Branchini \& Plionis, 1996). Therefore studies using clusters of galaxies as a 
probe, suggests that a significant part of the LG's motion is caused by 
structures at much greater distances than 100 h$^{-1}$ Mpc. Recently, this result
has been suported using X-ray cluster samples, which are free of the various
systematic effects from which the optical catalogues suffer (Plionis \& 
Kolokotronis, 1998). 
 
In the following sections, we will be discussing the dipole moment for the EMSS AGNs and the
ROSAT detected quasars sample, from which we will find the values of the bias parameters. 
Furthermore, to investigate the clustering of objects at small scales, we study their 
angular and spatial correlation properties. 

\noindent One of the most common statistical characterizations of the clustering properties of galaxies 
and other extragalactic objects is the three-dimensional two point correlation function 
(Peebles, 1980), $\xi(r)$, in which one uses information on redshift, longitude and latitude 
for each object. Note that this is the Fourier transform of the power spectrum of the galaxy 
distribution (Peacock \& Nicholson, 1991), where the power spectrum is the square of the
absolute value of the Fourier transform of the density field of galaxies. 
An easier way to investigate these clustering properties, however, is the 
two-dimensional two point angular correlation function, $w(\theta)$, 
in which we only need information on the longitude and latitude for each object.
 
\noindent Later on, we will be discussing the results of the auto and cross-correlation functions
between various samples, from which we can deduce the values of the relative bias parameters.

\section{Description of Data}

The EMSS is a statistically complete (96\%) and well-defined sample of 835 
serendipitous X-ray sources (of which more than 400, are AGNs) detected in IPC
images of the high Galactic latitude sky obtained with the IPC on board the 
Einstain Observatory. It has limiting sensitivities from 5(10$^{-14}$) to 
3(10$^{-12}$) erg cm$^{-2}$ sec$^{-1}$ in the 
0.3-3.5 keV energy band (typically $> 3.25$ (10$^{-14}$) erg cm$^{-2}$ sec$^{-1}$ in the 
0.5-2.0 keV energy band), redshifts up to 2.83 and a sky coverage of 
778 deg$^{2}$ (for further details, see Stocke et al.,1991; Gioia et al.,1990; 
Maccacaro et al., 1991).
 
The sample of 654 ROSAT detected radio quasars (Brinkmann et al., 1997), is 
an all sky survey with a uniform flux limit of a few (10$^{-13}$) erg cm$^{-2}$ sec$^{-1}$ 
(0.1-2.4 keV) and redshifts up to 3.886. This sample has been compiled from the 
Veron-Cetty-Veron catalogue detected by ROSAT in the ROSAT All-sky survey 
(Voges, 1992), as targets of pointed observations, or as serendipitous sources
from pointed observations as available publicly from the ROSAT point source 
catalogue (ROSAT-SRC, Voges et al., 1995)

\section{A Brief Review of the tests for Cosmological Evolution 
for Xray Selected AGNs}

To test the evolution in cosmological objects, one normally investigates the
logN-logS relation, the $< V_e / V_a >$ test and the differential luminosity 
function.
 
The logN-logS relation for the EMSS AGNs and the 654 ROSAT detected quasars are
shown in Figure 1. It appears that these AGNs have gone through a positive 
luminosity evolution with time, ie. they have either been more luminous or more
common $/$ numerous in the past. Also note that the flux at which the logN-logS
curve stops increasing is an approximation for the flux limit of the sample.
Errors have been estimated as pure Poissonian. Furthermore, using the 
maximum likelihood method, each curve has been fitted with a power law model 
. As can be seen, a steeper slope is found for the EMSS sample $(-1.57\pm0.1)$ 
compared to the 654 quasar sample $(-1.13\pm0.36)$, from which it might appear
that the EMSS AGNs have evolved more in time. However, considering the 1$\sigma$ 
errors, the slopes are consistent with -1.5 which is the value found for
non-evolving objects.
 
Next, the differential X-ray Luminosity Function (XLF) for both the EMSS sample and
the 654 quasar sample, are presented in Figure 2 and Figure 3. To obtain these, 
we have used the method of summing inverse accessed volumes, as 
discribed in Schmidt (1968); Maccacaro (1991) and Page et al. (1996). 
Therefore, the XLF is calculated as:

\begin{figure}
\centerline{\psfig{figure=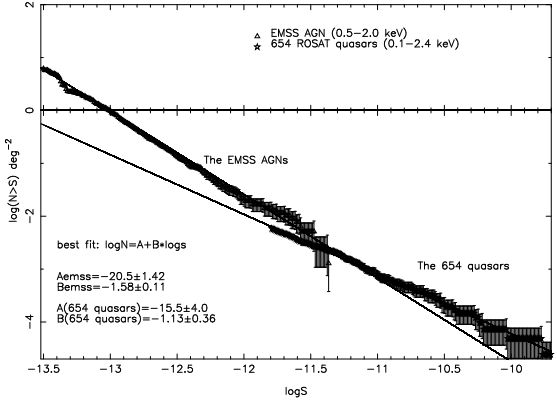,width=0.45\textwidth,angle=-90}}
\caption{The differential logN-logS relation for the EMSS and the 654 ROSAT detected AGNs}
\end{figure}

\begin{equation}
\Phi(L,z)=\frac {dN(L)}{dL} =
\sum_{1}^{N} \frac {1}{V_s.dL}                                 
\end{equation} 
 
\noindent where, $V_s$ is the volume in the universe searched to find a
particular object, and dL corresponds to the luminosity bins.

\noindent Also the error bars are calculated using: 
 
\begin{equation}
(\sum_{1}^{N} \frac {1}{V_s^2})^{1/2}
\end{equation}  

\begin{figure}
\centerline{\psfig{figure=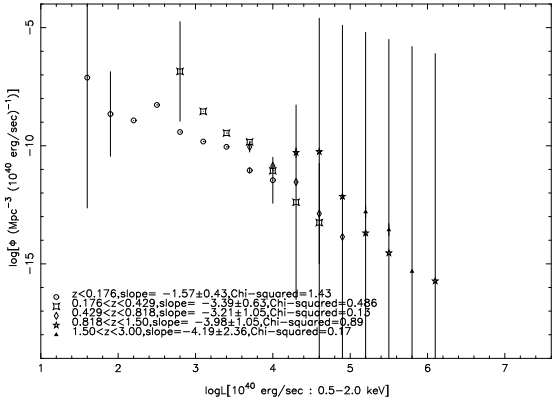,width=0.45\textwidth,angle=-90}}
\caption{The differential XLF for the EMSS AGNs}
\end{figure}
 
\begin{figure}
\centerline{\psfig{figure=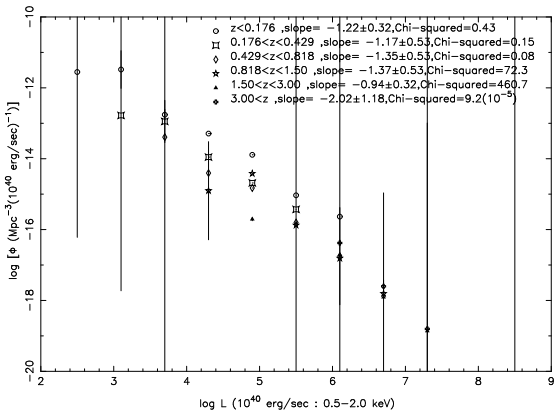,width=0.45\textwidth,angle=-90}}
\caption{The differential XLF for the 654 ROSAT detected quasars}
\end{figure}

\noindent Assuming only the luminosity evolution model (since this is more 
commonly accepted), a single power law model has been fitted to each individual
XLF corresponding to different redshift shells and a different slope has been 
found for each, as shown on the plots.

\noindent Note that it is more common to fit a LF with two power laws, but in this 
calculations, a single power law gave a reasonably good fit (The goodness of the 
fits can be tested using the 2D K-S test). Finding different slopes for each redshift shell, 
suggests that these AGNs have evolved differently at different periods of time.

\noindent The XLF is used to calculate the selection function, which is used in 
dipole and correlation function analysis. In dipole analysis, the weight 
corresponding to each object is proportional to the inverse of the selection 
function. Also in the analysis of the spatial correlation function, the random 
sample is selected so as to have the same selection function (ie. the
same redshift distribution) as the original sample. 
 
\noindent In most common cases of flux limited extragalactic object catalogues, one
has to take into account the effects of the consequent undersampling of the 
density field especially at large distances, where the radial selection
functions rapidly decline. Assuming that the unobserved galaxies are spatially 
correlated with those included in catalogue, the usual procedure to correct for
the missing population is to weight each observed object at a distance r by
a factor $\propto$ $\frac {1}{\Phi(r)}$, which is the reciprocal of the portion of the LF that can 
not be sampled at that distance due to the flux limit of the catalogue. Therefore,
the weight becomes: 
 
\begin{equation}
w_i=\phi^{-1} (r_i) r_i^{-2}
\end{equation}
 
\noindent where:
 
\begin{equation}
\phi(r) = \frac {1}{n} \int_{L_{min(r)}}^{L_{max}} \Phi(L)dL   
\end{equation}
 
\noindent is the selection function.
 
\noindent $\Phi(L)$ is the luminosity function, $L_{min(r)} = 4 \pi
r^{2} S_{lim}$ is the  
minimum luminosity that an extragalactic object can have in order to be visible
at a distance r which is determined by the flux limit of the sample. $L_{max}$ is 
the maximum luminosity of such an object.
 
\noindent Note that the average density, n, is calculated as:
 
\begin{equation}
n= \int_{L_{min}}^{L_{max}} \Phi(L) dL    
\end{equation}
 
\noindent where $L_{min}$ is the minimum luminosity of the object.

\noindent Note that the selection function is inversly proportional to r, and this acts as a compensating
effect for the fact that poor sampling occurs at large redshifts.
 
\section{The Cosmological Dipole Moment}

In this section, we give a summary on how to calculate the amplitude of the dipole 
growth curve as a function of the radial distance. Basically, the dipole moment is 
defined as:

\begin{equation}
D = \sum_{i=1}^{i=N} w_i r_i
\end{equation}

\noindent where $r_i$ is the unit vector pointing at the position of each object and N is
the total number of such objects within the distance considered. The weight 
$(w_i)$ is calculated as shown in equation 3. 

Using the linear perturbation theory, and the above definition of 
the dipole, the relation between the observed peculiar velocity of an observer, Vp,
and that predicted by its gravitational acceleration, g, as long as the two
vectors are well aligned is:

\begin{equation}
V_p(r) = \frac {H_o \beta}{4 \pi n} \sum_{i=1}^{i=N} \frac {\phi^{-1}(r_i) r_i}{|r_i|^3} 
= \beta g(r)
\end{equation}

\noindent where: $\beta =\frac {\Omega_o ^{0.6}} {b}$, (b is the bias parameter).

\noindent Note that $V_p(r)$ decreases with increasing r, therefore the more distant objects have
less contribution to the acceleration of the LG, but do continue to add to the peculiar velocity.

\noindent From the above equation, the three components of the accelaration generated by
each object on the LG can be calculated $(V_x,V_y,V_z)$. The force can then be 
smoothed out to guarantee linearity, where the function Smooth is defined as:

\begin{equation}
Smooth (r^2,r_{min}) = \frac {1}{max (r^2.r,r_{min}^3)}
\end{equation}

\noindent ie. each component is multiplied by the function smooth.

\noindent We then sum over shells to build up the three cartesian components of the
cumulative dipole (acceleration) at radius r. Finally we calculate the
amplitude of the cumulative dipole at the same radius, using:

\begin{equation}
V_{cum} = \sqrt (V_{cum-x}^2+V_{cum-y}^2+V_{cum-z}^2)
\end{equation}

\noindent where $V_{cum-x}$, $V_{cum-y}$, $V_{cum-z}$ are the three components of the 
cumulative acceleration.

\noindent Note that the moments of the objects are affected by the discretness effects or
the shot noise errors, which increase with redshift bacause of the rapidly
declining selection function. These effects introduce a variance in the dipole
amplitude and direction. To calculate these effects, we can use the method of 
Strauss et al. (1992), Hudson (1993) and Branchini \& Plionis (1996), as follows:

\begin{equation}
\sigma_z^2 = \sum_{i=1}^{i=N} w_i^2 (z_i / |r_i|^3)^2
\end{equation}

\noindent Therefore:

\begin{equation}
\sigma_{3D} = \sqrt (\sigma_x^2 +\sigma_y^2+\sigma_z^2)
\end{equation}

\noindent and the corresponding error along the line of sight (1D error) is:

\begin{equation}
\sigma_{1D} =\frac {\sigma_{3D}} {3^{1/2}}
\end{equation}

\section{The Results of the Dipole Moment Analysis}

The resulting dipole growth curves for the EMSS sample 
is shown in figure 4.

\begin{figure}
\centerline{\psfig{figure=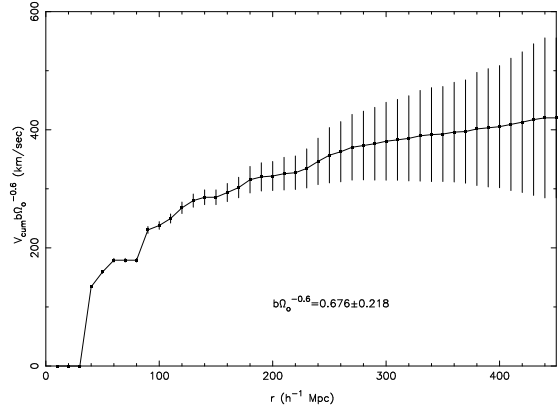,width=0.45\textwidth,angle=-90}}
\caption{The Dipole Growth Curve for the EMSS AGNs}
\end{figure}

As can be seen, there is no signal observed at small depths $(ie. < 40 h^{-1} Mpc)$,
indicating insignificant contribution to the LG motion from the nearby objects. This is 
expected, since there are not many objects in the EMSS sample with distances 
$< 40 h^{-1} Mpc$. The amplitude of the dipole growth curve saturates at rather
large depths of $> 300 h^{-1} Mpc$. However from distances of about $300 h^{-1}$ Mpc, 
the amplitude of the dipole increases only by small amounts. Therefore
$R_{conv} \sim 300-400 h^{-1} Mpc$, which is the depth from 
which these AGNs contribute to the acceleration of the LG via their 
gravitational attraction. Note that there are relatively large error bars, 
particularly at larger depths, which put limits to the accuracy of the value
of $R_{conv}$.

\noindent From this analysis, the value of $b_{EMSS} \Omega_o^{-0.6}$ is found to be $0.68 \pm 0.22$.
This depends on the present value of the cosmological density parameter (cf. $\Omega_o$ = 1
for the Einstein-de Sitter universe). For IRAS galaxies, $R_{conv} \sim 40-100 h^{-1} Mpc$, 
therefore comparing this with the one for the EMSS sample, suggests 
a dependence between the maximum effective depth of the sample and the convergence radius,  
ie. the deeper the sample (in distances), the larger the value of
$R_{conv}$. 

\noindent To further compare the dipole moments of the EMSS AGNs with that of the IRAS 1.2 Jy 
galaxies, we have selected a sample of 352 IRAS galaxies with the same 
selection function as the EMSS sample (ie. the same redshift distribution) and 
plotted the cumulative dipole growth curves for both samples, as shown in figure 5.

\begin{figure}
\centerline{\psfig{figure=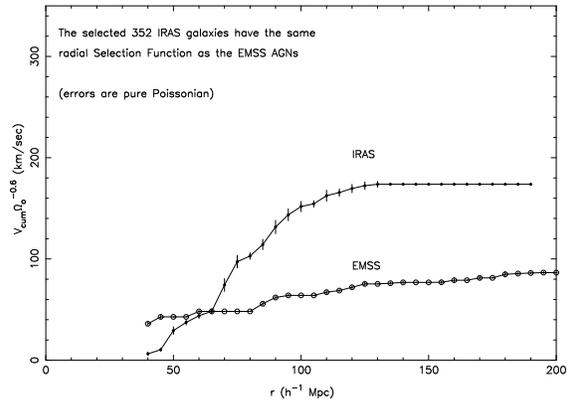,width=0.45\textwidth,angle=-90}}
\caption{Comparing the Dipole Growth curves for the EMSS AGNs and the
IRAS galaxies}
\end{figure}

\noindent In these analysis, we have plotted both curves 
with bias parameter of 1.0. As can be seen at almost all distances, the dipole moment of IRAS galaxies
dominates that of the EMSS. It appears that IRAS galaxies have stronger 
contributions to the motion of the LG than the EMSS AGNs. In other words, IRAS
galaxies are more clustered than the EMSS AGNs at scales of $< 200 h^{-1}$ Mpc (this
is compared to the underlying mass distribution), which is consistent with the values
of the bias parameter (assuming the Einstein-de Sitter universe). 

\noindent Similarly, the dipole growth curve for the 654 ROSAT detected radio quasars, is shown in Figure 6.

\begin{figure}
\centerline{\psfig{figure=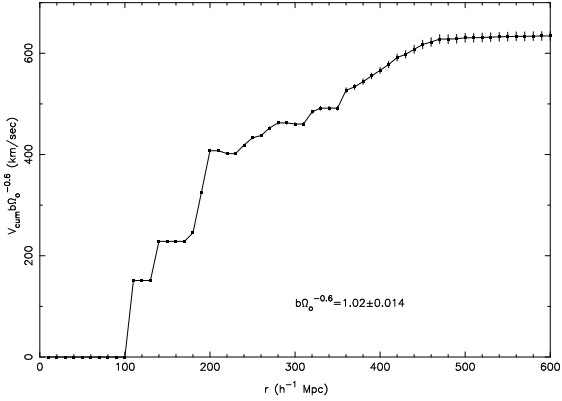,width=0.45\textwidth,angle=-90}}
\caption{The Dipole Growth Curve for the 654 ROSAT detected radio quasars}
\end{figure}

\noindent For this sample, $R_{conv} \sim 450-500 h^{-1} Mpc$, which suggests that 
these AGNs contribute to the LG motion from much deeper distances than the IRAS or the EMSS
samples. This can be explained as being due to the large redshifts of the 
objects in this sample. Also the amplitude at the saturation point, is found 
to be $\sim 617 \pm 8 km/sec$, from which we obtain $b_{654 quasars} \Omega_o^{-0.6}= 
1.02 \pm 0.01$, which suggests that these AGNs are less clustered than the 
IRAS 1.2 Jy galaxies, for which  $b_{IRAS} \Omega_o^{-0.6}= 
1.9 \pm 0.09$.

\noindent Finally, for a further investigations, we have plotted the dipole growth curves
for this sample and a sample of 178 IRAS 1.2Jy galaxies selected with the same selection 
function, as shown in figure 7.

\begin{figure}
\centerline{\psfig{figure=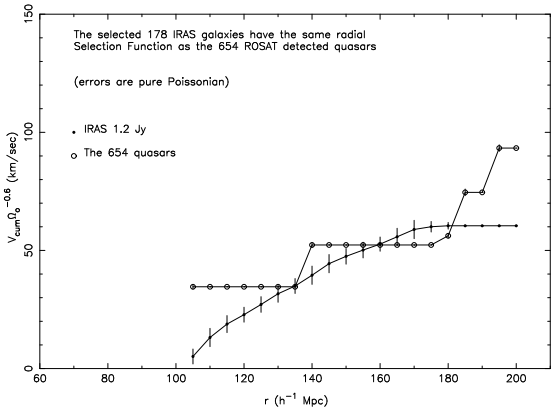,width=0.45\textwidth,angle=-90}}
\caption{Comparing the Dipole Growth curves for the 654 ROSAT detected quasars and the
IRAS galaxies}
\end{figure}

\noindent Note that in these, we have assumed the bias parameter of 1.0 for both curves. 
This has created some overlapping between the two curves, suggesting a similar degree 
of clustering for both samples.

\section {The Angular and Spatial Correlation Functions}
 
The cross-correlation function $\xi_{ab}$(r), between population $'a'$ and $'b'$, is defined
in terms of the joint probability $\delta P$ that a population $'a'$ object is in an 
infinitesimal volume element $\delta V_a$ and a population $'b'$ in $\delta V_b$ 
separated from $\delta V_a$ and $\delta V_b$ by $r_{ab}$:
 
\begin{equation}
\delta P = <n_a> <n_b> [1+\xi_{ab}(r_{ab})] \delta V_a \delta V_b
\end{equation} 

\noindent where $<n_a>$ and $<n_b>$ are the mean number densities of population $'a'$ and 
$'b'$ respectively. 

\noindent If galaxies were distributed randomly in the universe, $\xi(r)$ 
would be zero. If there is any distance scale over which objects tend to 
cluster, $\xi(r)> 0$, and if there is a scale at which the objects avoid each
other (anti-clustering), $\xi(r)$ is negative within that scale. $\xi(r)$ is calculated
as:

\begin{equation} 
\xi(r) = \frac {N_{DD}(r)}{N_{DR}(r)}\quad . \quad 
\frac {n_R}{n_D}  -1
\end{equation} 

\noindent where $N_{DD}$ and $N_{DR}$ refer to the number of data-data and data-random pairs, 
respectively; $n_D$ and $n_R$ refer to the mean densities of the real and random 
catalogues. 

\noindent Note that for the analysis of the spatial correlation function, the 
population of the random sample is chosen to be at least 50 times more than the 
real sample. This random sample is selected so as to have
the same selection function (ie. the same redshift distribution) as the real 
sample.
 
\noindent This function has been calculated for galaxies (Davis \& Peebles, 1983; Fisher 
et al., 1994; etc.), clusters of galaxies (Bahcall et al., 1984) and QSOs (
Shanks et al., 1987).
 
\noindent Note that the errors are calculated as the weighted Poissonian:

\begin{equation} 
\delta \xi(r) \sim  \sqrt \frac {1+\xi(r)}{N_{DR}} \sim \frac
{\sqrt N_{DD}}{N_{DR}} 
\end{equation} 

\noindent A common feature of the spatial correlation functions, is that they can be 
approximated by a power law model (on small scales, $r<10$ h$^{-1}$ Mpc, where the 
galaxy distribution is characterized by strong nonlinear clustering) 
of the form:
 
\begin{equation}
\xi(r)=(r/r_o)^{-\gamma}
\end{equation}

\noindent where $r_o$ and $\gamma$ are the correlation length and the power law index 
respectively. For galaxies, $r_o$ is of the order 5.4 h$^{-1}$ Mpc (Davis \& Peebles,
1983) with $\gamma \sim 1.7-1.8$ . This correlation length has been found to increase
as the scale of the system increases (ie. a larger value of $r_o$ is found for 
rich clusters and even a larger value is found for superclusters). Bahcall \&
Soneira (1983) found $r_o \sim 25$ h$^{-1}$ Mpc and $\gamma \sim 1.7$ for Abell clusters.
 
\noindent To find the best power law fit, we have used the method of $\chi^{2}$ 
minimisation. and the errors on each parameter correspond to $\delta \chi^{2}=1$.
Note that this method is valid only if the errors of the correlation function at
a fixed separation are Gaussian distributed and if the values at different
separations are uncorrelated. The power law model is usually fitted to the data 
for r $\leq$ 20 h$^{-1}$ Mpc and since these separations are typically much smaller than 
the size of the sample, the central limit theorem ensures that the distribution 
of errors will be Gaussian. 

\noindent According to the linear biasing theory (Kaiser, 1984), the two point 
correlation function between populations $'a'$ and $'b'$ can be approximated by:

\begin{equation} 
\xi_{ab}(r) = b_a b_b \xi_m(r)
\end{equation} 

\noindent where $b_a$ and $b_b$ are the bias parameters (ie. the contrast enhancement factor of the
distribution of tracers compared to the underlying mass distribution) for populations $'a'$ and $'b'$ 
respectively and $\xi_m(r)$ is the correlation function of the underlying mass density. 

\noindent b is defined as:

\begin{equation}
b=\frac{(\frac{\delta\rho}{\rho})_{tracer}}{(\frac{\delta\rho}{\rho})_{mass}}
\end{equation}

\noindent Calculating the correlation function of various tracers indicates 
the relative bias parameters of these tracers (Lahav, Nemiroff et al., 1990). Also, comparing
a cross-correlation function $(\xi_{IX}(r))$ with an auto-correlation function 
$(\xi_{II}(r))$, gives extra relative biasing between two types of mass tracers:
 
\begin{equation}
\frac {\xi_{IX}(r)}{\xi_{II}(r)}= \frac {b_X}{b_I}
\end{equation} 

\noindent Having described the calculation of the spatial correlation function, let
us now have a brief review of the angular correlation function. The angular correlation 
function can be expressed in terms of probability in a 
similar way as the spatial correlation function. If one considers two 
differential elements of solid angle on the sky, d$\Omega_1$ and d$\Omega_2$, then 
the joint probality, dP, that galaxies will occupy the two elements with an
angular separation of $\theta$ can be written as:
 
\begin{equation}
dP = n^{2} (1+w(\theta)) d\Omega_1 d\Omega_2
\end{equation} 

\noindent where n is the mean surface density of galaxies. Note that a random (Poisson)
distribution of galaxies yields w($\theta$) = 0 for all $\theta$. Also:
 
\begin{equation}
w= \frac {N_{DD}} {N_{RR}} -1
\end{equation} 

\noindent or:
 
\begin{equation}
w= \frac {N_{DD}}{N_{DR}} -1
\end{equation} 

\noindent Again, the errors are calculated as Poissonian errors. Finally, w($\theta$) can 
be approximated by a power law of the form:
 
\begin{equation}
w(\theta) = A \theta^{-\delta}
\end{equation} 

\noindent As before the best fit, can be found using the methods of $\chi^{2}$ 
minimisation, and the errors on each parameter correspond to $\delta \chi^2$=1

\section{The Results of the Angular and Spatial Correlation Function Calculations}
 
The results of the various angular and spatial auto and cross correlation
functions are shown in figures 8-15.

\begin{figure}
\centerline{\psfig{figure=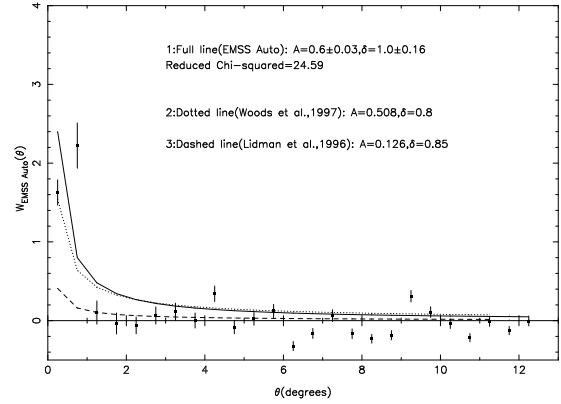,width=0.45\textwidth,angle=-90}}
\caption{The EMSS AGNs Angular Auto-Correlation Function}
\end{figure}

\begin{figure}
\centerline{\psfig{figure=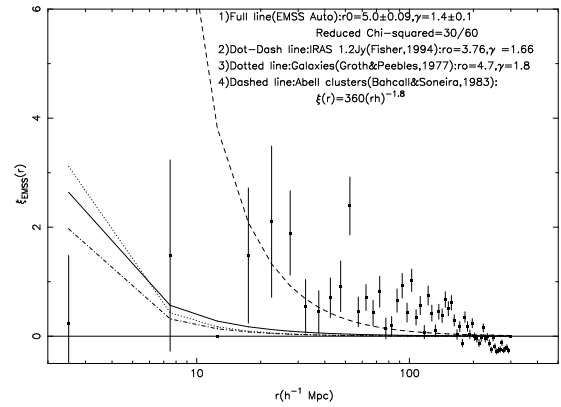,width=0.45\textwidth,angle=-90}}
\caption{The EMSS AGNs Spatial Auto-Correlation Function}
\end{figure}

\begin{figure}
\centerline{\psfig{figure=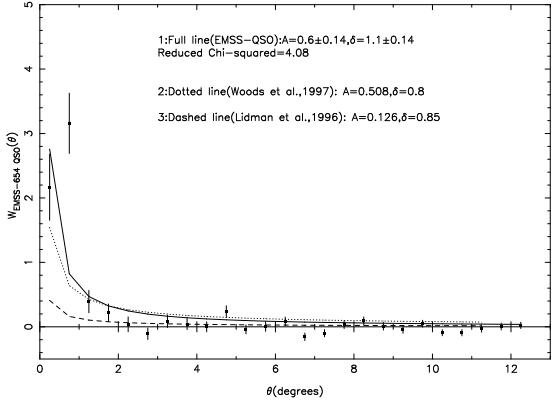,width=0.45\textwidth,angle=-90}}
\caption{The EMSS AGNs-654 QSO Angular Cross-Correlation Function}
\end{figure}

\begin{figure}
\centerline{\psfig{figure=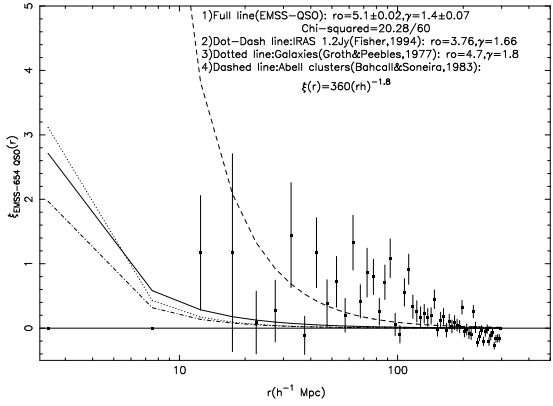,width=0.45\textwidth,angle=-90}}
\caption{The EMSS AGNs-654 QSO Spatial Cross-Correlation Function}
\end{figure}

\begin{figure}
\centerline{\psfig{figure=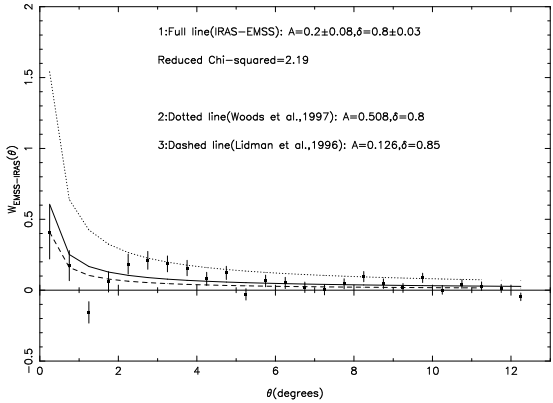,width=0.45\textwidth,angle=-90}}
\caption{The EMSS AGNs-IRAS 1.2 Jy galaxies Angular Cross-Correlation Function}
\end{figure}

\begin{figure}
\centerline{\psfig{figure=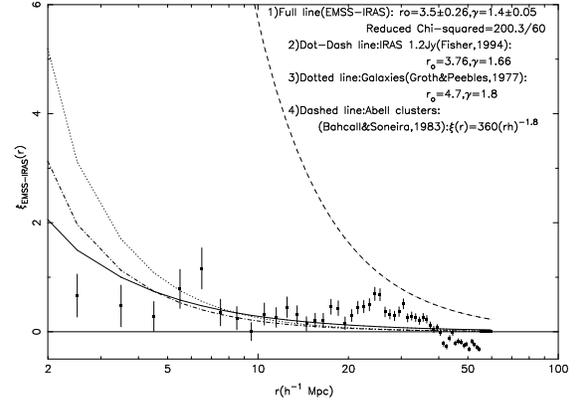,width=0.45\textwidth,angle=-90}}
\caption{The EMSS AGNs-IRAS 1.2 Jy galaxies Spatial Cross-Correlation Function}
\end{figure}

\begin{figure}
\centerline{\psfig{figure=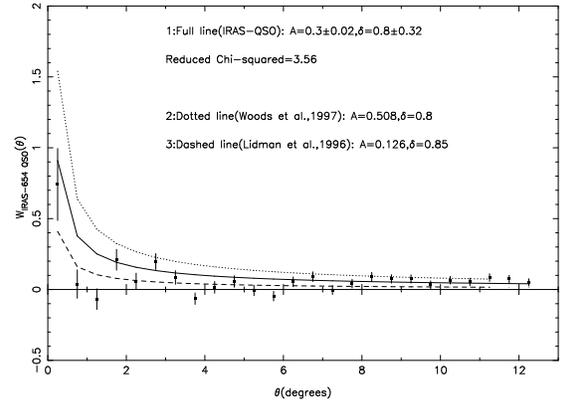,width=0.45\textwidth,angle=-90}}
\caption{The IRAS 1.2 Jy sample-654 QSO Angular Cross-Correlation Function}
\end{figure}
 
\begin{figure}
\centerline{\psfig{figure=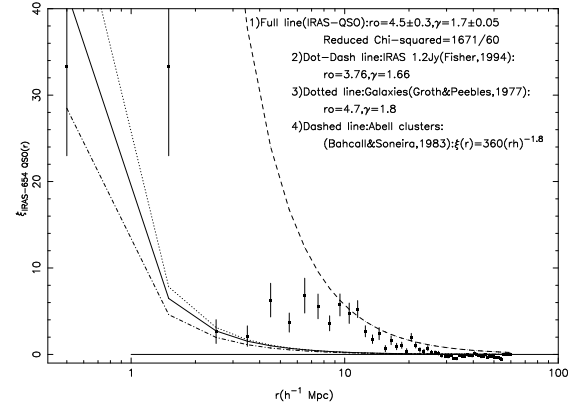,width=0.45\textwidth,angle=-90}}
\caption{The IRAS 1.2 Jy sample-654 QSO Spatial Cross-Correlation Function}
\end{figure}

The best power law fit to all the data analysed in this work are shown in solid lines.
For further comparisons, we have also plotted power law fits from previous work.
For angular correlation functions, these are: the fit found by Woods et al. 
(1997), studying the angular auto-correlation function for faint optical galaxies in 
high galactic latitude fields: $w(\theta)$ = 0.508 $\theta^{-0.8}$ 
(shown as the dotted line); and the fit by Lidman et al. (1996), studying the 
two point angular auto-correlation function for a sample of optical galaxies selected 
from two large format CCD surveys: $w(\theta)$ = 0.126 $\theta^{-0.85}$ 
(shown as the dashed line). 
 
\noindent For the spatial correlation functions, these are: The power law fitted by Fisher et 
al. (1994), calculating the spatial auto-correlation function for the IRAS 1.2 Jy galaxies 
(shown as the dot-dash curve): $\xi(r) = (r / 3.76)^{-1.66}$;
a general fit found for the galaxy auto-correlation (Groth \& Peebles, 1977), 
shown by the dotted line: $\xi(r) = (r / 4.7)^{-1.8}$; and finally the fit found for the 
Abell clusters auto-correlation (Bahcall \& Soneira, 1983), presented as the dashed line: 
$\xi(r)= 360(rh)^{-1.8}$.

\noindent It is important to note that there is a linear dependence between the scale of the system and
the correlation length (ie. the scale of clustering). $r_o$ is found to be $\sim 5 h^{-1} Mpc$ for
galaxies and $\sim 25 h^{-1} Mpc$ for Abell clusters. 
 
\noindent Furthermore, we have compared the average amplitudes for various samples for
different angular and distance separations, as shown in Tables 1-4.
 
\noindent Note that from all these comparisons, it appears that AGN / quasar samples are
less clustered than the galaxy samples, and this is consistent with the results
of the dipole analysis. Note that AGNs are rare distance objects, and this puts
limits to the accuracy of the correlation signals at small scales. In the
spatial correlation functions, there are relatively large error bars at small
scales $(r<10 h^{-1} Mpc)$ and the data is consistent with zero.
 
\noindent Finally, the relative bias parameters between various samples (at r = 12.5
h$^{-1}$ Mpc) are found to be:

\begin{equation} 
b_{IRAS} / b_{654 quasar}= 3.6 \pm 2.7
\end{equation}

\begin{equation}
b_{IRAS}/b_{EMSS}= 8.5 \pm 6.4
\end{equation}

\begin{equation}
b_{654 quasars} / b_{EMSS}= 2.4 \pm 0.2
\end{equation} 

\noindent which again suggests that AGNs / quasars are less clustered than the IRAS galaxies.

\begin{table*}
\footnotesize
\begin{center}
\begin{tabular}{cccccc}
 
$\theta$ &  $<W_{EMSS}>$ & $<W_{EMSS-654 QSO(Cross)}>$ & $<W_{IRAS}>$ & $<W_{Woods}>$ & 
$<W_{Lidman}>$ \\
 
$1.5^o<\theta< 5^o$ & $0.05\pm0.07$ & $0.074\pm0.09$ & $0.14\pm0.07$ & $0.21$ & $0.05$ \\
 
\end{tabular}
\end{center}
\caption{Comparing the average amplitude of the X-ray AGNs Angular Correlation functions with
those of IR/Optical galaxy Auto-Correlation functions}
\normalsize
\end{table*}

\begin{table*}
\footnotesize
\begin{center}
\begin{tabular}{cccccc}
 
$\theta$ &  $<W_{EMSS-IRAS(Cross)}>$ & $<W_{IRAS-654 QSO(Cross)}>$ & $<W_{IRAS}>$ & $<W_{Woods}>$ & 
$<W_{Lidman}>$ \\
 
$\theta< 1.5^o$ &$0.14\pm0.12$ & $0.23\pm0.13$ & $0.25\pm0.11$ & $0.79$ & $0.23$ \\

$1.5^o<\theta< 5^o$ & $0.14\pm0.06$ & $0.05\pm0.07$ & $0.14\pm0.07$ & $0.21$ & $0.05$ \\
 
\end{tabular}
\end{center}
\caption{Comparing the average amplitude of the AGNs-galaxies Angular Cross-Correlation functions 
with those of IR/Optical galaxy Auto-Correlation functions}
\normalsize
\end{table*}

\begin{table*}
\footnotesize
\begin{center}
\begin{tabular}{cccccc}
 
$r(h^{-1} Mpc)$ &  $<\xi_{EMSS}>$ & $<\xi_{EMSS-654 QSO(Cross)}>$ & $<\xi_{IRAS}>$ & $<\xi_{galaxies}>$ & 
$<\xi_{Abell clusters}>$ \\
 
$r<10$ &$0.86\pm1.49$ & no signals! & $1.15$ & $1.78$ & $39.4$ \\
 
$10<r<20$ & $0.81\pm0.6$ & $1.84\pm1.2$ & $0.11$ & $0.13$ & $2.96$ \\
 
$20<r<50$ & $7.4\pm0.64$ & $6.9\pm0.5$ & $0.03$ & $0.03$ & $0.7$ \\

\end{tabular}
\end{center}
\caption{Comparing the average amplitude of the X-ray AGNs Spatial Correlation functions with
those of IR/Optical galaxy Auto-Correlation functions}
\normalsize
\end{table*}

\begin{table*}
\footnotesize
\begin{center}
\begin{tabular}{cccccc}
 
$r(h^{-1} Mpc)$ &  $<\xi_{EMSS-IRAS(Cross)}>$ & $<\xi_{IRAS-654 QSO(Cross)}>$ & 
$<\xi_{IRAS}>$ & $<\xi_{galaxies}>$ & 
$<\xi_{Abell clusters}>$ \\
 
$r<10$ &$0.58\pm0.34$ & $10.3\pm3.2$ & $3.86$ & $7.25$ & $161.1$ \\
 
$10<r<20$ & $0.29\pm0.15$ & $2.15\pm0.58$ & $0.1$ & $0.13$ & $3.03$ \\

\end{tabular}
\end{center}
\caption{Comparing the average amplitude of the AGNs-galaxies Spatial Cross-Correlation functions 
with those of IR/Optical galaxy Auto-Correlation functions}
\normalsize
\end{table*}

\section{Discussion}

Having obtained different bias parameters for different samples of AGNs, using the dipole moment and
the standard methods of spatial and angular correlation functions, let us now briefly discuss the
calculation of the spatial correlation function for a fractal.

\noindent Analysis of galaxy and cluster correlations based on the concepts and methods of modern statistical 
Physics, led to the suggestion that galaxy correlations are fractals and not homogeneous to
the limits of the available catalogs (Pietronero, 1987; Coleman et al., 1992).

\noindent According to these analysis, galaxy structures are highly irregular and self-similar$:$ 
all the available data are consistent with each other and show fractal correlations 
(with dimensions, $D \sim 2$) up to the deepest scales probed so far (eg. 1000 $h^{-1} Mpc$). 
If galaxy distribution becomes really homogeneous at a scale $\lambda_o$ within the sample 
in question (ie. if $\lambda_o \ll R_{eff}$, where $R_{eff}$ is the effective depth of 
the sample), then one has $r_o$ = $\lambda_o$ 2$^{1/(D-3)}$. For a fractal of dimension 
$D \sim 2$, therefore$:$ $r_o \sim \lambda_o / 2$. If on the other hand, the fractal correlations 
extend up to the sample limits, then the resulting value of $r_o$ has nothing to do with the real 
properties of the galaxy distribution, but it is fixed by the size of the sample.

\noindent The expression of $\xi(r)$ in the case of fractal distribution is:

\begin{equation}
\xi(r)= \frac{3-\gamma}{3}(\frac{r}{R_s})^{-\gamma} -1
\end{equation}

\noindent where $R_s$ is the depth of the spherical volume where one computes the average density.

\noindent The so-called correlation length, $r_o$, is a linear function of the sample size, $R_s$,

\begin{equation}
r_o= (\frac{3-\gamma}{6})^{1/\gamma}R_s
\end{equation} 

\noindent and hence it is a spurious quantity without physical meaning, but 
it is simply related to the 
sample size. Note that according to the above equation, there is a linear dependence between 
the $scale/size$ of the system and the correlation length , and this is consistent with the 
results from the standard power law model of the correlation function. In other words, 
as the size of the sample increases $(ie. R_s)$, the value of the corrrelation length increases 
(cf. $r_o \sim 5 h^{-1} Mpc$ for galaxies, and $r_o \sim 25 h^{-1} Mpc$ for Abell clusters).

\noindent Finally note that if the distribution is fractal, even the dipole moment is affected, and its 
value depends on even very distant objects, although weighted by $1/r^2$.

\section{Summary of Conclusions}
 
From the above analysis (the dipole moment and the correlation function), 
it appears that the EMSS AGNs are less clustered than the sample of 654 ROSAT detected 
quasars which itself is less clustered than the IRAS galaxies. This was numerically 
shown by the values of the bias parameters.
 
\noindent From the dipole analysis, we deduced the values of
$R_{conv}\sim 300-400 h^{-1} Mpc$ 
for the EMSS AGNs and $\sim 450-500 h ^{-1} Mpc$ 
for the quasar sample, from which it
appears that these AGNs / quasars have a gravitational effect on the LG motion from much
deeper distances than IRAS galaxies $(R_{conv} \sim40-100 h^{-1} Mpc)$ 
and Abell clusters $(R_{conv} \sim 160 h^{-1} Mpc)$. 
Also the amplitudes of the dipole growth 
curves at the saturation region are found to be $~ 405.4 \pm 103.36$
km/sec for  
the EMSS AGNs and $~ 617.3 \pm 8.35 km/sec$ for the quasar sample, from which we find: 
$b_{EMSS} \Omega_o^{-0.6}=0.68 \pm 0.22$ for the EMSS sample, and 
$b_{654 quasars} \Omega_o^{-0.6}$ = $1.02 \pm 0.014$ for the quasar sample. Comparing the 
maximum amplitudes at the saturation region, suggests that the more sensitive the sample
(ie. the lower the flux limit), the less clustered it is compared to the underlying mass
distribution.

\noindent Assuming that $b \Omega_o^{-0.6}\sim 1$, which is supported by deep AGN samples, we 
conclude $\Omega_o \simeq 1$ if $b\simeq 1$. However to obtain $\Omega_o \simeq 0.3$ (Krauss, 1998), 
requires b significantly less than 1.
 
\noindent From the correlation function analysis, it appears that the AGN / quasar samples are 
less clustered than the galaxies, which is consistent with the results of the dipole 
calculations. The values of the relative bias parameters were found to be: 
$b_{IRAS} / b_{654-quasar}= 3.61 \pm 2.7$, $b_{IRAS}/b_{EMSS}= 8.46 \pm 6.4$ and
$b_{654-quasars} / b_{EMSS}= 2.35 \pm 0.2$. However the validity of these values is subject
to the critisism that they are sample depth dependent.

\end{document}